\documentclass[useAMS,usenatbib]{mn2e}
\usepackage{journals}
\usepackage{graphicx}
\usepackage{natbib}

\title[The binary companion of PSR\,J1740$-$3052]{The binary companion of PSR\,J1740$-$3052}

\author[Bassa et al.]  {C.\,G.\,Bassa$^{1}$\thanks{email:
    bassa@jb.man.ac.uk}, W.\,F.\,Brisken$^2$, G.\,Nelemans$^3$, I.\,H.\,Stairs$^4$, B.\,W.\,Stappers$^1$, M.\,Kramer$^{5,1}$\\
$^1$Jodrell Bank Centre for Astrophysics, The University of Manchester, Manchester, M13\,9PL, United Kingdom\\
$^2$National Radio Astronomy Observatory, Socorro, NM 87801, USA\\
$^3$Department of Astrophysics, IMAPP, Radboud University Nijmegen, Toernooiveld 1, 6525\,ED, Nijmegen, The Netherlands\\
$^4$Department of Physics and Astronomy, University of British Columbia, 6224 Agricultural Road, Vancouver BC V6T\,1Z1, Canada\\
$^5$Max Planck Institut f\"ur Radioastronomie, Auf dem H\"ugel 69, 53121 Bonn, Germany
}

\begin{document}

\date{Accepted 2010 December 20. Received 2010 December 15; in original form 2010 November 12}

\pagerange{\pageref{firstpage}--\pageref{lastpage}} \pubyear{2002}

\maketitle

\label{firstpage}

\begin{abstract}
  We report on the identification of a near-infrared counterpart to
  the massive ($>11$\,M$_\odot$) binary companion of pulsar
  J1740$-$3052. An accurate celestial position of PSR\,J1740$-$3052 is
  determined from interferometric radio observations. Adaptive optics
  corrected near-infrared imaging observations show a counterpart at
  the interferometric position of the pulsar. The counterpart has
  $K_\mathrm{s}=15.87\pm0.10$ and $J-K_\mathrm{s}>0.83$. Based on
  distance and absorption estimates from models of the Galactic
  electron and dust distributions these observed magnitudes are
  consistent with those of a main-sequence star as the binary
  companion. We argue that this counterpart is the binary companion to
  PSR\,J1740$-$3052 and thus rule out a stellar mass black hole as the
  pulsar companion.
\end{abstract}

\begin{keywords}
  binaries: general -- stars: early-type -- pulsars: general --
  pulsars: individual: PSR\,J1740$-$3052 -- infrared: stars --
  astrometry
\end{keywords}

\section{Introduction}
Among the radio pulsars that are found in binary systems, four belong
to a subgroup where the pulsar has the characteristics of a young,
isolated pulsar, but is orbiting a massive companion
($\ga3$\,M$_\odot$) in a rather wide (50--2000\,day orbital period)
and eccentric ($e>0.5$) orbit \citep{sta04,kbjj05}. The companion to
the first of these systems, PSR\,B1259$-$63 \citep{jml+92} was
identified as a $V=10$ early-type B2e star. PSR\,J0045$-$7319, located
in the SMC, is the second system and has a $V=16$ B1\,V companion
\citep{kjb+94}. No information on a possible counterpart has been
published for the fourth system, PSR\,J1638$-$4725 \citep{lfl+06}.

The third system, PSR J1740$-$3052, was discovered by \citet{sml+01}
in the Parkes Multibeam Survey \citep{mlc+01} as a young (350\,kyr)
pulsar. Subsequent timing observations showed that the pulsar is part
of a binary system with an orbital period of 231\,d and an
eccentricity of 0.57. The mass function of the pulsar yielded a lower
limit to the mass of the companion of 11\,M$_\odot$. A candidate
counterpart to PSR\,J1740$-$3052 was identified in near-infrared
observations. From $K$-band spectra, the counterpart was classified as
late-type, having a spectral type between K5 and M3
\citep{sml+01}. This identification is at odds with the binary
parameters which set the Roche lobe radius significantly smaller than
that expected for a late-type star at the estimated
distance. Subsequent phase-resolved spectroscopic observations by
\citet{tsw+10} confirmed the unlikeliness of the star being the binary
companion, as no significant radial velocity variations consistent
with the expected binary orbit were found.

The nature of the massive binary companion remains unclear, and could
be either an early-type main-sequence star or a compact
remnant. Though \citet{sml+01} reported small variations in the
dispersion measure of PSR\,J1740$-$3052 near periastron, which would
be consistent with a stellar companion, they could not conclusively
rule out a stellar mass black hole as the binary companion.

In this paper we report on interferometric radio observations to
obtain an accurate position of the PSR\,J1740$-$3052 and adaptive
optics corrected near-infrared observations to search for a stellar
counterpart to PSR\,J1740$-$3052. Section\,2 describes the
interferometric observations, while Section\,3 details the astrometry
and photometry of the near-infrared imaging. We discuss the results in
Section\,4 and conclude in Section\,5.

\section{Radio observations}
The NRAO Very Large Array (VLA) was used to determine the position of
the pulsar under project code AS769. The observation was made over a
5\,h period starting at UTC 23:19 on September 3rd, 2003 using the VLA
in its most extended A-configuration. This observation included the
Pie Town antenna for even greater East-West resolution.  Spectral line
mode mode 2AC was used to provide 25\,MHz of bandwidth centered at
1678.4\,MHz on each circular polarization. The correlator integration
time was set to 5\,s. The observation consisted of a 600\,s
observation of 3C286 followed by 14 loops containing three sources:
90\,s on PMN\,J1751$-$2524, 90\,s on PMN\,J1820$-$2528 and 930\,s on
PSR\,J1740$-$3052.

Data reduction was performed within AIPS in a standard manner.  Flux
density and bandpass calibration were performed on 3C286.  Calibrator
source PMN\,J1751$-$2524 was used as the position reference.  Its
assumed position was taken to be
$\alpha_\mathrm{J2000}=17^\mathrm{h}51^\mathrm{m}51\fs2630$,
$\delta_\mathrm{J2000}=-25\degr24\arcmin00\farcs060$ \citep{fma+04}.
Phases and amplitudes determined on this source were applied to
PMN\,J1820$-$2528 and PSR\,J1740$-$3052.  Imaging was performed using
the robust weighting factor of unity as a compromise between
resolution and sensitivity; the resultant beam size was
$0\farcs91\times2\farcs17$ oriented $2\degr$ East of North.
Wide-field imaging of the PSR\,J1740$-$3052 data was performed in
order to clean side-lobes from numerous bright sources over the entire
$25\arcmin$ field of view.  Most notably, the $5\arcmin$ diameter
object known as \textit{The Tornado} (see \citealt{gfs+03} and
references within), is only $8\farcm1$ Southwest of the pulsar. Due to
the lack of short interferometer spacings, the ability to model and
clean this source well was limited.  Fortunately the pulsar lies
outside the worst side-lobes which emanate mostly North and South from
the object. The pulsar was detected less than one beam-width from the
position reported by \citet{sml+01} with a flux density of
$0.26\pm0.04$\,mJy. Despite the best effort to clean the side-lobes of
confusing sources, the pulsar was found on top of residual large scale
structure. In order to fit the position of the pulsar, linear flux
density slices were made along the right ascension and declination
axes through the center of the detected pulsar. Each of these two
slices were then fit for a 1-D Gaussian and an underlying linear
gradient. The resultant best fit position of the pulsar is
$\alpha_\mathrm{J2000}=17^\mathrm{h}40^\mathrm{m}50\fs001$,
$\delta_\mathrm{J2000}=-30\degr52\arcmin04\farcs30$. The $1\sigma$
uncertainties on this position are $0\farcs069$ in right ascension and
$0\farcs21$ in declination. The effect of an incorrectly modeled
underlying gradient was found to result in deflections much smaller
than the fit uncertainty so are not considered. The final uncertainty
reported above is entirely due to measurement error.

The 0.8\,Jy VLA calibrator source PMN\,J1820$-$2528 has nearly the
same angular separation ($6\fdg6$) from the calibrator source
PMN\,J1751$-$2524 as does PSR\,J1740$-$3052 ($6\fdg0$) and was used as
an astrometric check source. It was imaged using the same parameters
as used for PSR\,J1740$-$3052. Its fit position
($\alpha_\mathrm{J2000}=18^\mathrm{h}20^\mathrm{m}57\fs84909$,
$\delta_\mathrm{J2000}=-25\degr28\arcmin12\farcs5403$) differed from
its cataloged position
($\alpha_\mathrm{J2000}=18^\mathrm{h}20^\mathrm{m}57\fs8486$,
$\delta_\mathrm{J2000}=-25\degr28\arcmin12\farcs584$;\,\citealt{fma+04})
in each axis by an amount much smaller than the statistical
uncertainty of the pulsar position.

\section{Infrared Observations}
Adaptive optics corrected near-infrared imaging of the field
containing PSR\,J1740$-$3052 was obtained with NACO (NAOS-CONICA;
\citealt{lhb+03,rlp+03}) at the ESO Very Large Telescope on Paranal,
Chile in July and August of 2006. A log of the observations is given
in Table\,\ref{tab:log}. Images were obtained in the $J$, $H$ and
$K_\mathrm{s}$-bands using the S27 camera, which provides a
1k$\times$1k chip with a $0\farcs027$\,pix$^{-1}$ pixel scale yielding
a $28\arcsec\times28\arcsec$ field-of-view. The late-type candidate
counterpart of \citet{sml+01} was used as a natural guide star for the
wavefront sensor which corrects the atmospheric distortions in the
incoming wavefront. The \texttt{N90C10} dichroic was used for
$K_\mathrm{s}$ observations, while the \texttt{K} dichroic was used
for $J$ and $H$-band imaging.  The dithered NACO observations were
used to create a sky frame containing the contribution of both the
dark current and the inhomogeneities in the sky for each filter and
exposure time combination. After subtraction of this sky frame, the
science frames were registered using integer pixel offsets and
averaged.

Our aim was to obtain an accurate absolute astrometric calibration of
the NACO images using the USNO CCD Astrograph Catalog (UCAC3;
\citealt{zfg+10}) to allow comparison with the interferometric pulsar
position. The positional uncertainties of the standards in this
catalog are 20\,mas for stars with $R\la14$ and 70\,mas for $R\la16$,
but with 1.3 stars per square arc-minute the star density in the
direction of PSR\,J1740$-$3052 is low and none of the standards
coincides with the NACO field-of-view.

In order to allow calibration against the UCAC3 catalog, imaging
observations were obtained with two further instruments. Narrow-band
near-infrared images at 1.19 and 2.25\,$\mu$m were taken with ISAAC
\citep{mcb+98}, also at the VLT in June 2006. The total exposure times
were 936\,s and 800\,s respectively. This instrument has a
1k$\times$1k CCD with a $0\farcs148$\,pix$^{-1}$ pixel scale with a
$2\farcm5\times2\farcm5$ field-of-view. The observations were
corrected for dark current and flat-field effects and registered and
averaged using standard routines. Finally, $R$, $I$ and $z$-band
images were obtained with the Wide Field Imager (WFI;
\citealt{bmi+99}) at the 2.2\,m ESO telescope on La Silla, Chile in
August 2006. This instrument is a mosaic of 8 2k$\times$4k CCDs each
with a pixel size of $0\farcs238$\,pix$^{-1}$ and a
$8\arcmin\times16\arcmin$ field-of-view. We only use the data from the
CCD that contains the pulsar location. Bias-subtraction and
flat-fielding were preformed using standard routines. The $I$ and
$z$-band images were further corrected for fringe frames which were
constructed from the dithered science observations. All data reduction
was done using the Munich Image Data Analysis System (MIDAS).

Stars on an $8\arcmin\times8\arcmin$ subsection of a 45\,s $R$-band
WFI image were compared to UCAC3 standards. A total of 79 astrometric
standards coincided with the image, of which 72 were not saturated and
appeared stellar and unblended. An astrometric solution was determined
by fitting for position and a four parameter transformation matrix. By
iteratively removing outliers the solution converged using 67
standards providing rms residuals of $0\farcs073$ in right ascension
and $0.070$ in declination.

This astrometric solution was transferred from the WFI image to the
NACO images by using the 1.19\,$\mu$m and 2.25\,$\mu$m ISAAC images to
generate tertiary and quarternary astrometric catalogs. Pixel
positions on the ISAAC images were corrected for geometric distortion
using the May 2005 transformation provided by
ESO\footnote{http://www.eso.org/sci/facilities/paranal/\\instruments/isaac/inst/field\_distortion.html}.
The transfer of the astrometric solution from the $R$-band WFI image
to the 1.19\,$\mu$m ISAAC image used 56 stars, providing rms residuals
of $0\farcs022$ in right ascension and $0\farcs023$ in
declination. The solution was then transferred to the 2.25\,$\mu$m
image, using 120 stars yielding rms residuals of $0\farcs012$ in right
ascension and $0\farcs011$ in declination. Finally, 29 stars were used
to transfer the solution to the NACO $K_\mathrm{s}$-band image obtained on
July\,13. The residuals of the transfer were $0\farcs012$ in right
ascension and $0\farcs016$ in declination.

The total uncertainty in the astrometry of the NACO image is the
quadratic sum of the uncertainties in the astrometric solutions, which
amount to $0\farcs078$ in right ascension and $0\farcs076$ in
declination and is dominated by the uncertainties in the calibration
of the WFI image by the UCAC3 catalog. Even though multiple transfers
of the astrometric solution are needed, it is still more accurate than
the direct calibration of the 2.25\,$\mu$m ISAAC image using 160 stars
from the 2MASS catalog \citep{scs+06}, which yields residuals of
$0\farcs11$ in both coordinates.

\begin{figure}
  \centering
  \includegraphics[width=8cm]{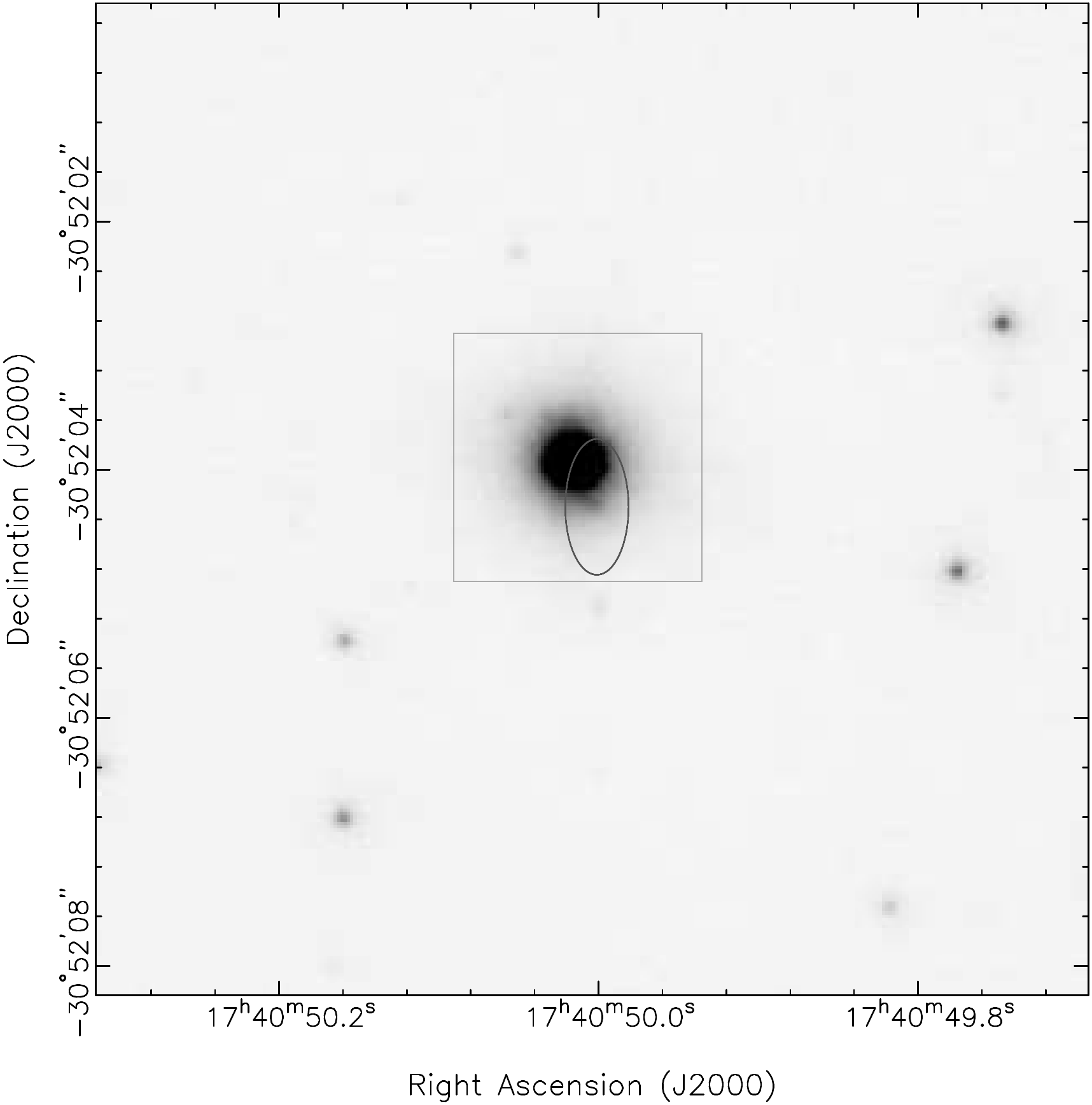}
  \caption{A $8\arcsec\times8\arcsec$ subsection of the
    $K_\mathrm{s}$-band images obtained with NACO on July\,13th,
    2006. The ellipse denotes the 95\% confidence error region of the
    interferometric radio position of the pulsar. The bright star
    near the center is the late-type star which \citet{tsw+10} showed
    is unlikely to be the binary companion to PSR\,J1740$-$3052. This
    star is located on the edge of the error region. Only a fainter
    star is present in the error ellipse. The $2\arcsec\times2\arcsec$
    box is the field shown in Fig.\,\ref{fig:fig2}.}
  \label{fig:fig1}
\end{figure}

\begin{figure}
  \centering
  \includegraphics[width=8cm]{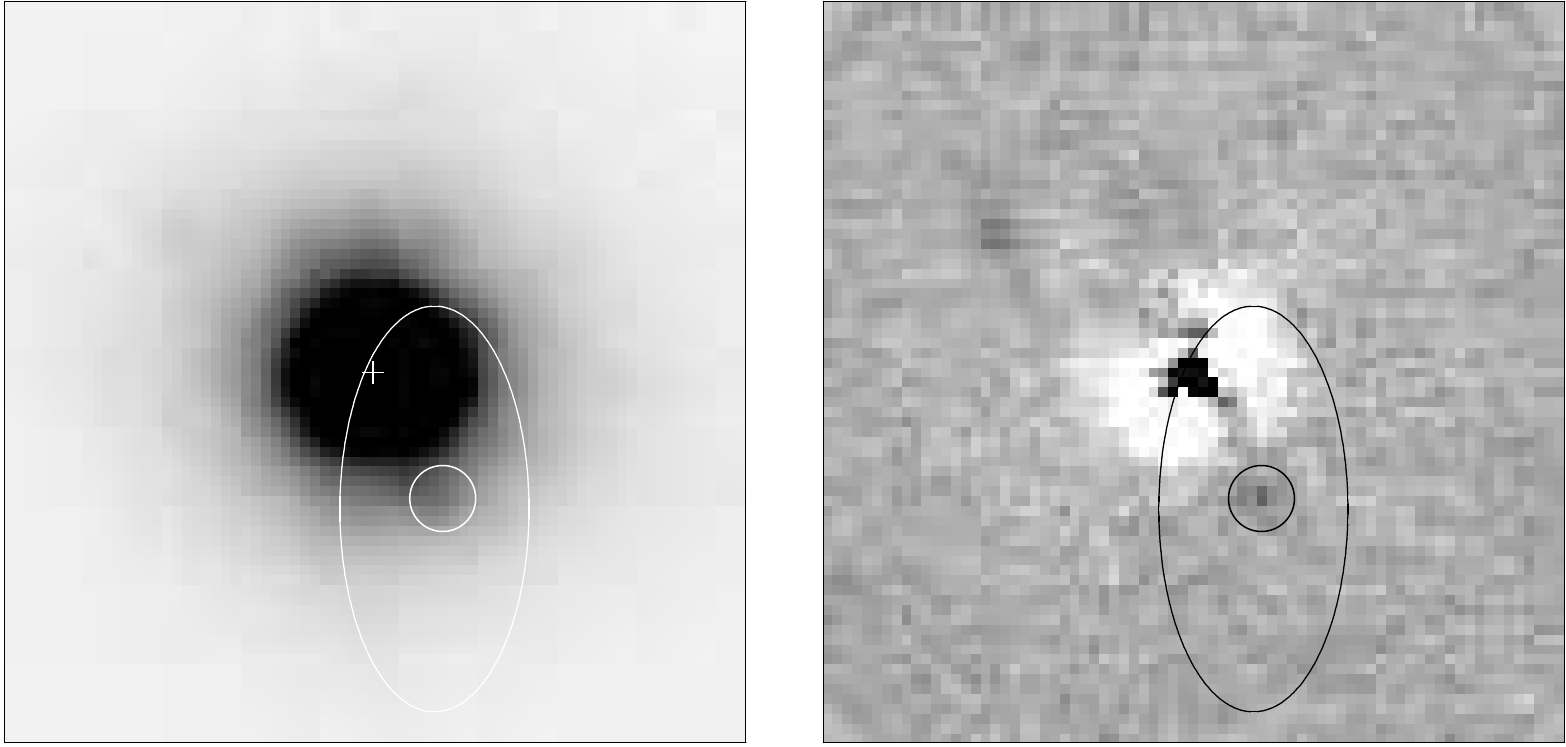}
  \caption{A $2\arcsec\times2\arcsec$ subsection of
    Fig.\,\ref{fig:fig1} centered on the bright late-type star,
    showing the original image and with the stellar profile of the
    late-type star subtracted. Some artifacts from the late-type star
    are still present due to imperfect definition of the
    point-spread-function. The ellipse denotes the 95\% confidence
    error region on the interferometric position of
    PSR\,J1740$-$3052. The counterpart to PSR\,J1740$-$3052 is
    encircled in both panels.}
  \label{fig:fig2}
\end{figure}

Combined with the uncertainty of the interferometric position of
PSR\,J1740$-$3052 the 95\% confidence region on the NACO images is an
ellipse with a semi-minor axis of $0\farcs26$ in right ascension and a
semi-major axis of $0\farcs55$ in declination. This confidence region
is shown in Fig.\,\ref{fig:fig1} and is located $0\farcs18$ West and
$0\farcs37$ South of the late-type star, which itself is located just
inside the 95\% confidence region. Inspection of the NACO images shows
that the July 13th, 2006 $K_\mathrm{s}$-band image contains a star
located inside the error ellipse, see Fig.\,\ref{fig:fig2}. The
position of the counterpart in that image is
$\alpha_\mathrm{J2000}=17^\mathrm{h}40^\mathrm{m}50\fs001\pm0\fs006$
and $\delta_\mathrm{J2000} =-30\degr52\arcmin04\farcs27\pm0\farcs08$,
which is offset from the interferometric position of PSR\,J1740$-$3052
by $0\farcs00\pm0\farcs10$ in right ascension and
$0\farcs03\pm0\farcs22$ in declination. Based on the stellar density
of the $K_\mathrm{s}$-band image, we estimate that the probability of
finding a random star this close to the interferometric position of
PSR\,J1740$-$3052 is only 0.05\,per\,cent.

\begin{table}
  \footnotesize \centering
  \caption[]{NACO Observation log. The columns give the date and time
    of the observations, the integration time $t_\mathrm{int}$, the
    filter used, the observed FWHM $\sigma$ of the stellar profiles
    and the observed magnitude or limit on the magnitude in this
    filter used.}\label{tab:log} \renewcommand{\footnoterule}{}
  \begin{tabular}{l@{\hspace{0.4cm}}
      l@{\hspace{0.3cm}}
      r@{\hspace{0.3cm}}
      c@{\hspace{0.3cm}}
      c@{\hspace{0.3cm}}
      r@{\hspace{0.3cm}}
    }
    \hline
    \hline
    \multicolumn{2}{l}{Date \& Time (UT)} & $t_\mathrm{int}$ (s) & Filter & $\sigma$ & Mag. \\
    \hline
    July\,13 & 04:11--05:03 & $195\times10$ & $K_\mathrm{s}$ & $0\farcs13$ & 15.87 \\
    & 05:43--05:52 & $50\times5$ & $H$ & $0\farcs21$ & $>15.1$ \\
    & 05:55--06:57 & $47\times50$ & $J$ & $0\farcs36$ & $>16.7$ \\[0.2ex]
    August\,14 & 00:03--00:50 & $31\times50$ & $K_\mathrm{s}$ & $0\farcs25$ & $>12.8$ \\[0.2ex]
    August\,21 & 00:13--00:22 & $50\times5$ & $H$ & $0\farcs25$ & $>14.1$ \\
    & 00:25--01:26 & $240\times10$ & $J$ & $0\farcs39$ & $>16.1$ \\
    & 01:29--02:00 & $48\times25$ & $K_\mathrm{s}$ & $0\farcs20$ & $>13.5$ \\
    \hline
  \end{tabular}
\end{table}

Point-spread-function (PSF) photometry was performed on the averaged
NACO images using DAOphot\,II \citep{ste87} running within MIDAS. An
analytical Moffat profile \citep{mof69} with a fixed exponent of
$\beta=2.5$ combined with an empirical look-up table was used to
represent the PSF. The parameters of the PSF were kept constant over
the image. Due to the low apparent stellar density on the NACO images,
the majority of the stars on the images were used to determine the
PSF.

The instrumental $J\!H\!K_\mathrm{s}$ PSF magnitudes were calibrated
against values from the 2MASS catalog \citep{scs+06}. Due to the small
field-of-view of NACO, 4 stars were used to determine zero-point
magnitude offsets. No color terms were fitted for. The rms residuals
of the fits were 0.09\,mag in $J$-band and 0.06\,mag in both $H$ and
$K_\mathrm{s}$.

The counterpart to PSR\,J1740$-$3052 is detected in the July 13th
$K_\mathrm{s}$-band image, where it has $K_\mathrm{s}=15.87\pm0.10$.
Limits on the magnitude of the counterpart in the other images were
determined using simulations. For each image a set of copies were
generated where an artificial star of a certain magnitude was placed
on the position of the counterpart. Each of these copies was then
photometered using the same routines as the original. The $3\sigma$
detection limit was set when the artificial star was recovered with a
signal-to-noise of 3. The limiting magnitudes are given in
Table\,\ref{tab:log}. The most stringent limits set $J>16.7$ and
$H>15.1$. 

\section{Discussion}
We will now use the $K_\mathrm{s}$-band detection and $J$ and $H$-band
non-detections of the counterpart and compare them with values
predicted for the companion of PSR\,J1740$-$3052.

\begin{figure}
  \centering
  \includegraphics[width=8cm]{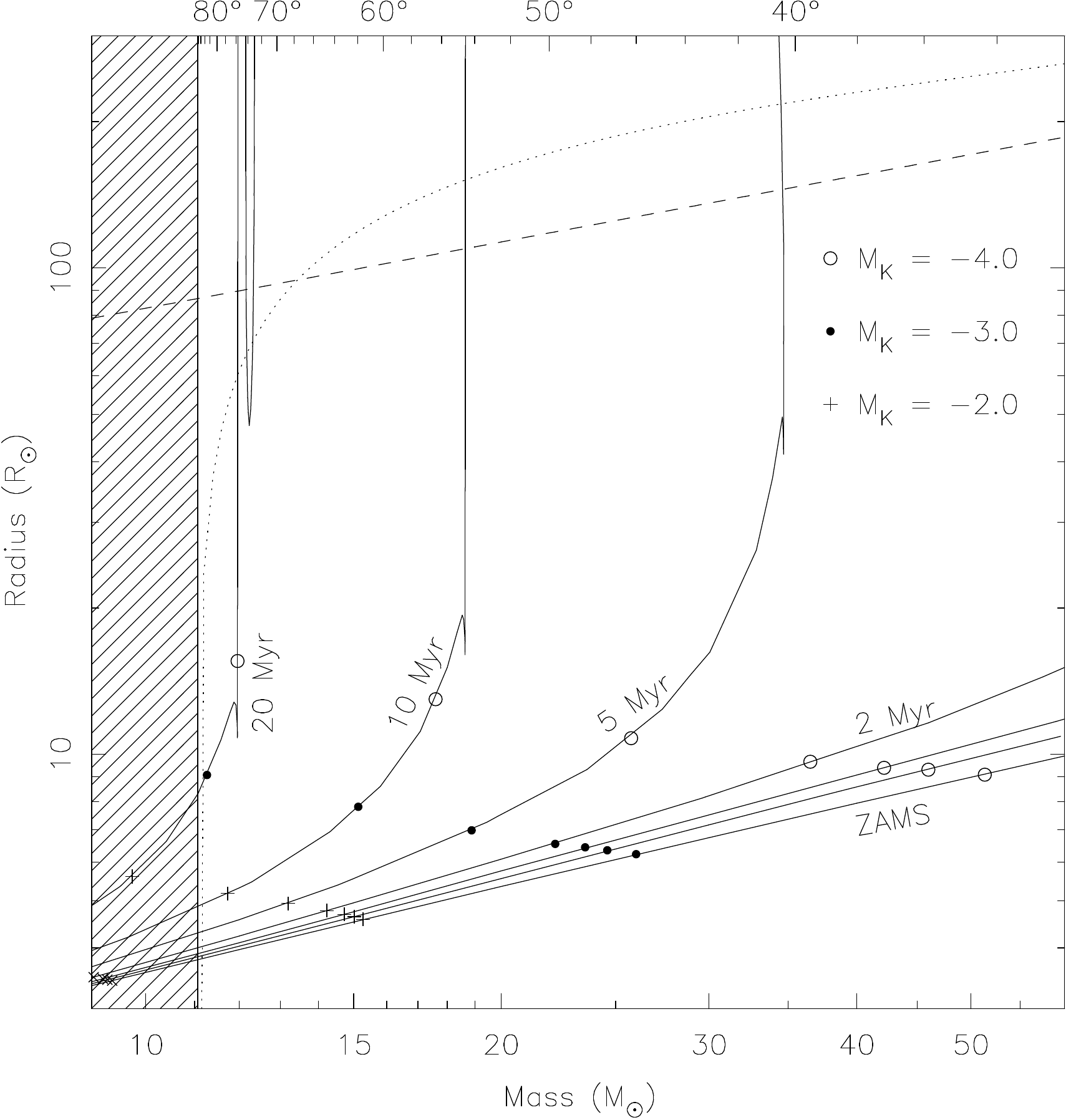}
  \caption{Constraints on the properties of the binary companion of
    PSR\,J1740$-$3052. Assuming a canonical pulsar mass of
    1.4\,M$_\odot$, the orbital parameters measured by \citet{sml+01}
    limit the companion to a mass of $M_\mathrm{c}>11$\,M$_\odot$.
    The Roche lobe radius at periastron is shown with the dashed line,
    setting an upper limit to the companion radius. A further upper
    limit, shown with the dotted line, on the companion radius is set
    by the absence of eclipses of the radio signal. Stars with radii
    larger than this limit would eclipse the pulsar when it passes the
    line-of-sight. Stellar isochrones from the Padova models by
    \citet{mgb+08} are shown with the solid lines, ranging from
    zero-age main-sequence models to models at ages of 0.5, 1, 2, 5,
    10 and 20\,Myr. Each isochrone shows the location where the star
    reaches absolute $K_\mathrm{s}$-band magnitudes of $M_k=-2$, $-3$
    and $-4$. The observed range of $1.2>M_K>-3.0$ shows that the
    companion is on the main-sequence.}
  \label{fig:fig3}
\end{figure}

The radio timing observations by \citet{sml+01} place several
constraints on the properties of the companion. Assuming the pulsar
has a canonical mass of 1.4\,M$_\odot$, the measurements of the
orbital period and semi-major axis limit the mass of the companion to
$M_\mathrm{c}>11$\,M$_\odot$. The radius of the companion is
constrained by the radius of the Roche lobe at the periastron distance
of the orbit. A further upper limit on the radius is set by the lack
of eclipses of the pulsar seen in the radio
observations. Figure\,\ref{fig:fig3} shows these constraints. Assuming
the magnetic field of the pulsar is a dipole, the rotational
properties of the pulsar can be used to determine the age of the
pulsar. For PSR\,J1740$-$3052 this characteristic age is 350\,kyr, and
binary evolution dictates that the companion must be older than that,
as the prior evolution of the pulsar progenitor must be involved
\citep{th06}.

At a Galactic longitude of $l=357\fdg81$ and latitude of $b=-0\fdg13$,
PSR\,J1740$-$3052 is located only $2\fdg2$ from the Galactic Center,
and hence the electron and dust column densities are substantial. At a
dispersion measure (DM) of 740.9\,cm$^{-3}$\,pc$^{-1}$ \citep{sml+01},
PSR\,J1740$-$3052 is one of the pulsars with the largest DM. For its
line-of-sight, distance estimates from models of the Galactic electron
density distribution based on the observed DM range from 8 to 14\,kpc
\citep{tc93} and 6.3 to 11\,kpc \citep{cl02}.

The \textsc{Cobe} dust maps by \citet{sfd98} set the maximum reddening
for this line-of-sight at $E_{B-V}=18$. Using the \citet{sfd98}
relative extinction coefficients, this is equivalent to a maximum $V$
and $K_\mathrm{s}$-band absorption of $A_V\la59$ and $A_K\la6.7$. The
nearest distinct line-of-sight of the Galactic extinction model by
\citet{mrr+06} is within $8\arcmin$ of PSR\,J1740$-$3052 and estimates
$K_\mathrm{s}$-band absorption of $A_K=1.9$ at a distance of 5.0\,kpc,
up to $A_K=2.5$ at 10.4\,kpc. This model determines the extinction
distribution by comparing 2MASS observations with predictions of a
Galactic stellar population synthesis model \citep{rrdp03}. For
comparison, the Galactic extinction model by \citet{dcl03}, where red
clump stars in the 2MASS observations are used to derive distance
dependent extinction, estimates $A_K$ of 1.2 at 5.0\,kpc, 1.7 at
10\,kpc and 3.0 at 15\,kpc. Hence, based on the observed
$K_\mathrm{s}$-band magnitude of the counterpart and a conservative
estimate on the range of distances ($d=5$ to 15\,kpc) and absorptions
($A_K=1.2$ to 3.0), the absolute $K_\mathrm{s}$-band magnitude of the
counterpart falls in the range of $M_\mathrm{K}=1.2$ to $-3.0$.

Stellar isochrones by \citet{mgb+08} show that main-sequence stars
have $M_J-M_K=(J-K_\mathrm{s})_0>-0.25$. Hence the observed limit on
$J-K_\mathrm{s}>0.83$ limits the $J-K_\mathrm{s}$ reddening to
$E_{J-K}>1.08$. Using the \citet{sfd98} extinction coefficients this
sets $A_K>0.74$, which is consistent with the estimates from Galactic
absorption models. The observed limit on $H-K_\mathrm{s}$ does not
provide a better constraint.

Figure\,\ref{fig:fig3} shows the predictions for mass and radius of
the \citet{mgb+08} isochrones. They show that zero-age main-sequence
stars with masses up to 25\,M$_\odot$ will have absolute
$K_\mathrm{s}$-band magnitudes that satisfy the observed range of
$M_\mathrm{K}=1.2$ to $-3.0$. Less massive but more evolved
main-sequence stars up to ages of 20\,Myr at 11\,M$_\odot$ also
satisfy this range. The constraints on the mass and brightness of the
companion of PSR\,J1740$-$3052 show that the companion must be on the
main-sequence. Furthermore, since the isochrones predict that
main-sequence stars more massive than 11\,M$_\odot$ have $M_K\la-1.5$
the $K_\mathrm{s}$-band distance modulus must be larger than
$(m-M)_K>17.4$, which suggests PSR\,J1740$-$3052 is located at the far
end of the distance ranges estimated from Galactic electron
distribution models.

\section{Conclusions}
Since its discovery, it was unclear if the binary companion of
PSR\,J1740$-$3052 was an early-type main-sequence star, a late-type
giant, or a stellar mass black hole, though the data favored a
main-sequence type companion. Using accurate astrometric radio
interferometry and adaptive optics corrected near-infrared imaging
observations, we show that the late-type star located near the pulsar
position is located at the edge of the 95\% confidence error ellipse
of the interferometric radio position of the pulsar. This further
strengthens the case made by \citet{sml+01} and \citet{tsw+10} that it
is not the binary companion to PSR\,J1740$-$3052.

In the wings of the late-type star, and near the center of the error
ellipse, we find a counterpart whose observed $K_\mathrm{s}$-band
magnitude is consistent with the expected mass and age of the binary
companion to PSR\,J1740$-$3052 at the estimated distance and reddening
to the system. We argue that this counterpart is the binary companion
to PSR\,J1740$-$3052. Our observations show that the companion is on
the main-sequence, which is consistent with the observations of both
the small variations in the dispersion measure near periastron of the
binary orbit, and the periastron advance being due to tidal and spin
quadrupoles \citep{sml+01}. These observations rule out the need for a
black hole or a giant as the binary companion.

At $K_\mathrm{s}=15.87$ the companion is within reach of near-infrared
spectrographs. Adaptive optics would be a necessity to resolve the
companion in the wings of the late-type star which is almost 6\,mag
brighter in $K_\mathrm{s}$. To extract the spectrum of the companion
an algorithm such as that of \citet{hyn02} would be needed to resolve
the blend. Spectra of the companion could be used to further confirm
the identification through radial velocity variations and the spectral
classification.

With a massive main-sequence companion, the PSR\,J1740$-$3052 system
likely experienced a previous phase of mass transfer in which the
progenitor of the pulsar lost its hydrogen envelope and the companion
accreted a significant amount of mass (for a review, see
\citealt{th06}). The subsequent (type Ib) supernova caused the orbit
to become eccentric. In the next few million years the companion will
evolve off the main-sequence and tidal forces will circularize the
orbit when the star approaches Roche lobe overflow. The ensuing common
envelope phase will likely lead to a merger and the formation of a
Thorne-Zytkow object \citep{tz75} as there is not
enough energy in the orbit to unbind the envelope of the subgiant
(e.g.\,\citealt{dt10}).

\section*{Acknowledgments}
The National Radio Astronomy Observatory (NRAO) is a facility of the
National Science Foundation operated under cooperative agreement by
Associated Universities, Inc. This research is based on observations
made with ESO Telescopes at the La Silla or Paranal Observatories
under program ID 077.D-0683. Pulsar research at UBC is supported by
an NSERC Discovery Grant.

\bibliographystyle{mn2e}

\label{lastpage}

\end{document}